# Mask mandates and COVID-19: A Re-analysis of the Boston school mask study

Tracy Beth Høeg, MD, PhD[1,2][†] Ambarish Chandra, PhD[3][†] Ram Duriseti, MD, PhD[4] Shamez Ladhani, MRCPH, PhD[5,6] and Vinay Prasad, MD, MPH[1]

[1]Department of Epidemiology and Biostatistics, University of California-San Francisco
[2]Department of Clinical Research, University of Southern Denmark, Odense, DK
[3]Department of Management, UTSC and Rotman School of Management, University of Toronto
[4]Department of Emergency Medicine, Stanford School of Medicine
[5]Immunisation and Countermeasures Division, UK Health Security Agency (UKHSA), London, UK.
[6]Paediatric Infectious Diseases Research Group, St. George's University of London, London, UK.

[†]These authors contributed equally

**Contact information**: Tracy Beth Høeg, Department of Epidemiology and Biostatistics, University of California-San Francisco, 550 16th Street, 2nd Floor, San Francisco, CA, 95148. tracy.hoeg@ucsf.edu. 415-476-2300.

**ABSTRACT**

**Background:** A recent epidemiological analysis of staggered policy implementation reported a 29.4% reduction in COVID-19 cases by maintaining school mask mandates in the greater Boston area during the first half of 2022. The robustness of their results and the appropriateness of methodology are explored.

**Methods:** Using data from the Massachusetts Department of Elementary and Secondary Education and the Centers for Disease Control and Prevention, we re-analyze differences in COVID-19 incidence in school districts that did and did not lift mask mandates using the same districts as the original study and expanded the analysis to the entire state of Massachusetts. We present changes in case rates and differences in prior immunity in areas with different mask lifting policies.

**Results:** The Boston and Chelsea districts, which maintained mask mandates, were outliers in terms of size, demographics, and testing. We failed to find a notable change in student cases in mask mandate districts compared with the 70 districts in the original study (-0.08/1000; p=0.98) and found a slight *increase* compared with a statewide control group +3.63/1000 (p=0.291). Results were similar for students and staff combined. Districts that dropped mask mandates first experienced the largest decreases in cases (22% drop vs 12% in the masked districts). There was a moderate to strong relationship ($R^2$ = 0.35-0.66; p-values <0.001) between prior community infection burden and district case rates in Spring 2022, with prior immunity alone explaining as much as two-thirds of the variation in case rates in Spring of 2022.

**Conclusions:** We fail to find any consistent notable negative relationship between school mask mandates and infection rates in the Greater Boston Area or state of Massachusetts during the 2021-2022 academic year.

**Introduction**

The use of face coverings for protection against SARS-CoV-2 infection has been highly controversial. Previously, pooled analysis of randomized population-level studies of medical and N95 masks to prevent the spread of respiratory viruses failed to find robust evidence of benefit.[1,2] Observational studies of mask mandate effectiveness in educational settings have had mixed findings.[3,4] In a recent, large epidemiological study, Cowger and colleagues analyzed the relationship between district masking policies and COVID-19 cases.[5] Using a specific Difference-in-Differences (DiD) methodology developed by Callaway and Sant'Anna[14], the authors assessed the impact of mask mandate removals on COVID-19 incidence at three timepoints starting February 28th, 2022. The authors concluded that, among school districts in the greater Boston area, lifting mask mandates resulted in 44.9 additional cases per 1000 students and staff or a 29.4% (95% CI, 21.4-37.5) increase.

Their findings, however, hinge on meeting DiD preconditions. The authors contend, without providing data, that before masking requirements were lifted, DiD estimates were "essentially zero" to support the assumption of parallel trends. The authors performed univariate sensitivity analyses on multiple time-varying covariates (community incidence, COVID-19 vaccination rates, and community test-positivity) and determined that the "parallel trends" assumption was not violated before statewide masking-mandate date were lifted.

DiD methodology should, however, only be used in the absence of time-varying confounders, otherwise the assumptions required for causal inference are not met. There were multiple time-varying confounders in Cowger et al including, prior infection-based immunity, number of vaccine doses, time since last vaccination dose, and SARS-CoV-2 testing practices, including changes in home/non-pooled testing rates (Cowger et al, Figures S7C, S7E and **Figure S2**).[5] With such critical variables changing over time, use of the DiD method is precluded even if the parallel trends assumption was met.

Additionally, school districts that impose mask mandates are also likely to impose other interventions (cohorting, physical distancing, staggered mealtimes, ventilation upgrades, etc), which will confound any analysis of mask mandates alone.[12] Consequently, the study by Cowger et al should be evaluated as any other observational study and include adjustments for confounding variables. We, therefore, re-analyzed Cowger *et al.*'s findings using multiple alternative methodologies and include a larger state-wide control group.

**Methods**

We use publicly-available data of district infection rates among staff in the 72 Boston area school districts studied by Cowger et al. We expanded the analysis to the entire state of Massachusetts,[6] and, unlike Cowger et al., did not exclude districts reporting no cases for >5-10 weeks. We incorporated Supplementary Data from the Centers of Disease Control and Prevention (CDC) to corroborate our findings.[7] We compared ratios and magnitudes of differences in case rates for the 2021/22 academic year (40 calendar weeks, September 1st to June 15, 2022) among students and staff before and after mask mandate removal. Though February 28th was the official date of the mask policy changes, we like Cowger, use Thursday March 3rd, 2022, as a proxy date of the mask policy change because it is the first case-reporting date. Boston-Chelsea were the only districts that maintained mask mandates.

**Intervention Analysis and Primary Outcome**

As in Cowger et al, the primary exposure was presence or absence of masking requirement in each reporting week. As in Cowger et al, a school district was considered to have lifted its masking requirement if the requirement had been lifted before the first day of the reporting week (Thursday). The primary outcome was SARS-CoV-2 incidence among students and staff, considered together and separately.

**Data Sources**

For each school district, data regarding weekly SARS-CoV-2 infections, student enrollment, and staffing during 2021–2022 were publicly available from the Massachusetts Department of Elementary and Secondary Education (MDESE) and supplemented with case surveillance data from the CDC.[6,7] Throughout the study period, MDESE required standardized weekly reporting of all positive tests for SARS-CoV-2 among students and staff, regardless of symptoms, testing type or program (e.g., testing of symptomatic persons or pooled polymerase-chain-reaction testing), and testing location (community-setting or school-setting).[22] We used March 3rd to June 15th, 2022, as the dates Boston-Chelsea were the treatment group (mask mandates) and the remaining 70 districts were controls.

**Extension of study period visualization**

We replicated Cowger et al's methodology to plot district case rates for students, staff and both for January 1st through June 15th, 2022, by mask mandate policy and date of mask mandate removal. We then extended these plots back to the start of the academic year.

**Extension to all districts state-wide**

The removal of mask mandates applied to the entire state but Cowger et al restricted their analysis to 72 school districts in the Greater Boston area without providing a rationale for limiting their data.[8] Several districts that technically lie outside greater Boston are geographically closer to Boston than other districts within the metropolitan area.

For consistency with Cowger et al, we restricted our analysis to districts that were not charter, vocational or technical schools, leaving 289 (of 399) school districts. We then searched for evidence of any districts statewide which retained mask mandates (**Supplementary Material**). While 5–7 districts had sporadic masking rules beyond March 17, the vast majority of the 289 districts did not.[8,9,10,11] We compared Boston-Chelsea with all school districts in the state.

**Statistical Analyses**

We calculated the proportions of infections pre- and post-March 3rd, 2022 for students and staff, separately and together, and performed non staggered DiD (**Supplementary Material**). We also calculated differences in case ratios before and after mask policy change by district mask-mandate lifting date. Raw data and code: https://github.com/tracybethhoeg/bostonmaskstudy-reanalysis

**Results**

Of the 72 Greater Boston Area school districts identified by Cowger at al, the study population consisted of 294,084 students and 46,530 staff. Boston-Chelsea, which retained mask mandates, had 52,243 students and 9323 staff. As in Cowger et al. analysis, after excluding 7 districts with no SARS-CoV-2 infections reported over >5-10 weeks, the 70 districts within the NECTA had 241,841 students and 37,206 staff. We identified 217 additional districts in MA (534,270 students, 82,280 staff) to analyze a total of 289 school districts across all Massachusetts (828,354 students, 128,809 staff).[6] Only two (Boston-Chelsea) mandated masking post-March 3rd, 2021.

**Figure 1** depicts total student enrollment against the fraction of the student population that is white for the 72 schools included in Cowger et al. The Boston school district is a notable outlier in terms of both size and racial composition. Boston's enrollment of 46,169 students is four times larger than the next largest districts (Newton: 11,974, Quincy: 9404). Since Chelsea's population is much smaller than Boston's, any differences between districts that ended versus maintained mask mandates is likely to be driven by results in Boston alone. Additionally, factors idiosyncratic to Boston – such as lower testing or reporting rates, or behavioral differences – would have disproportionate effects on aggregate outcomes versus other districts. Importantly, too, the fraction of non-white students was 85% in Boston and 94% in Chelsea, compared with 34% in the other 70 districts.[5] Consequently, using Boston schools in one arm of a DiD analysis was inappropriate because of size and population differences characteristics compared to other districts.

**Figure 1:** Student Enrollment and District Population

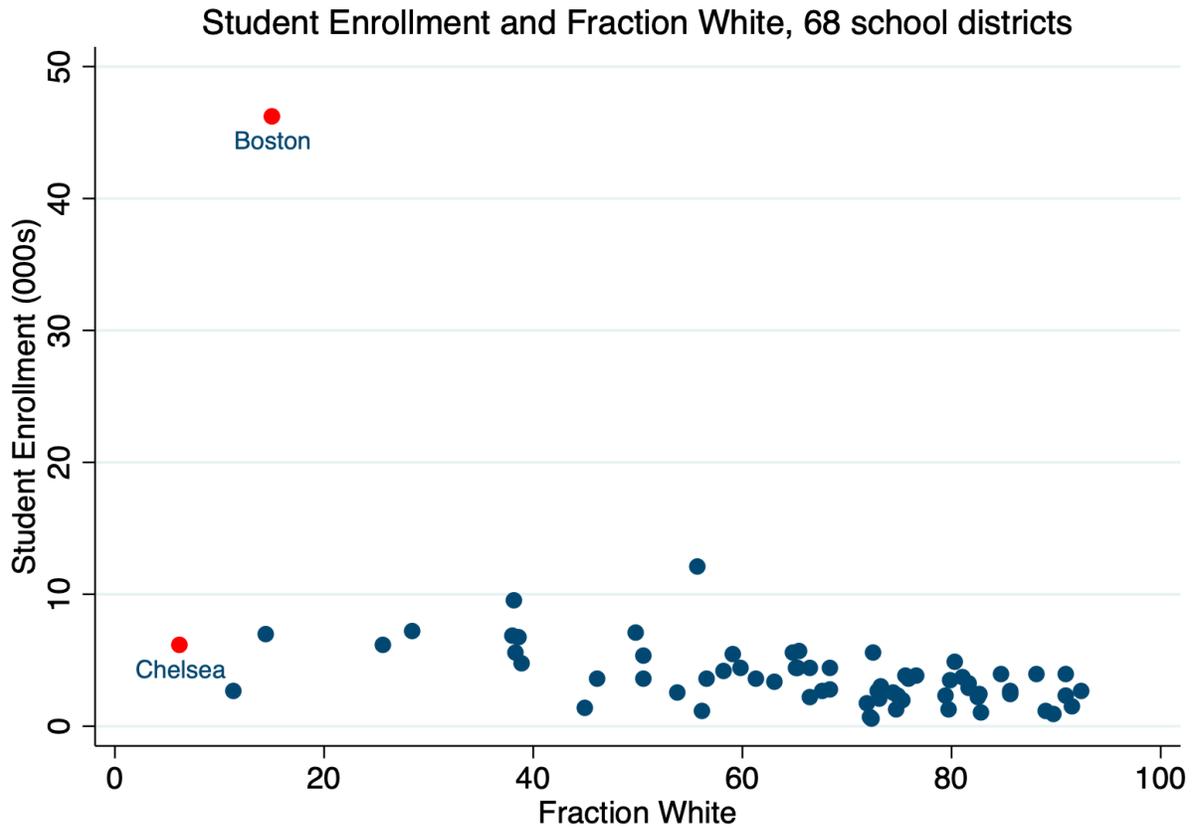

Legend: Total student enrollment by % white students in the district, colored by mask mandates status. Red=districts which retained mask mandates in Cowger et al. Blue=districts which dropped mask mandates between February 28th and March 14th, 2022.

**Extension of the period under study and to all districts state-wide**

**Figure 2a** replicates Figure 1B from Cowger et al, showing weekly SARS-CoV-2 infection rates per 1000 students during January–June 2022 in districts according to when the mask mandate was removed. Case rates were similar, with parallel trends, when mask mandates were removed, but their statistical analysis was performed for the entire academic year, so it is unclear why the time-period in this figure was restricted. We extended the time-period to the entire 2021-22 academic year and found that Boston-Chelsea consistently reported lower case rates than other districts, even during the first Omicron wave when statewide mask mandate was in place, indicating that factors other than mask mandates were more likely to explain the consistently lower infection rates in Boston-Chelsea during the study period.

**Figure 2a and b.** Replication of Figure 1B from Cowger et al.

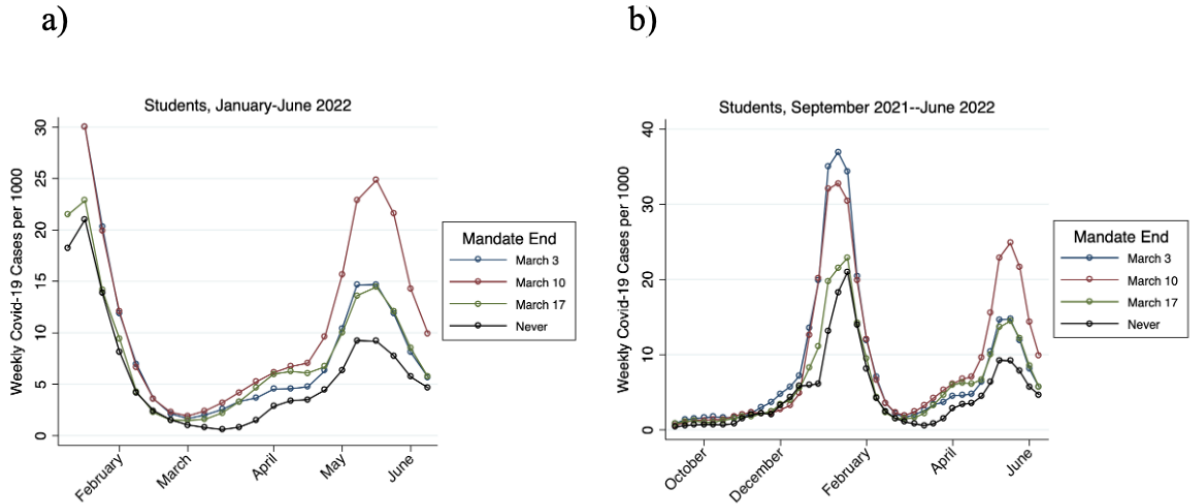

Legend: (a) show weekly Covid-19 cases per 1000 students for the period January–June 2022 and (b) extension of the time period of Figure 2a to the entire 2021-2022 academic year. District numbers are colored by mask mandate drop dates (or none shown in black)

Furthermore, the largest decline in cases (22%) was experienced by the districts which dropped their mask mandates first, which was greater than the 12% decrease in Boston-Chelsea (**Table 1**). The arrows show that districts which dropped mask mandates earlier had larger drops or smaller increases in cases, which is inconsistent with a causal relationship between changes in mask policies district case rates. This inverse dose-response relationship with unmasking does not support the use of the staggered Callaway Sant'Anna DiD analysis employed by Cowger et al.[14] We, therefore, provide a conventional DiD analysis that does not rely on the one-to-two-week differences in timing used by the staggered analysis.

**Table 1.** Case rate changes from pre- to post-March 3rd 2023 by multiple baseline time periods and intervention periods, stratified by district mandate lifting date or none. If none, results shown by different baseline periods based on the other districts' mandate end dates. Red arrows demonstrate larger drops in case rates with earlier mandate removal.

| Rates are Cases per 1000 individuals in the district | Mandate Removed 3/3 | Mandate Removed 3/10 | Mandate Removed 3/17 | Never Removed (3/3 cutoff date used) | Never Removed (3/10 cutoff date used) | Never Removed (3/17 cutoff date used) |
|---|---|---|---|---|---|---|
| Pre-Mandate Removal Average from Sep 16 '21 | 9.0 | 7.9 | 5.5 | 4.9 | 4.7 | 4.5 |
| Pre-Mandate Removal Average from Dec 2 '21 | 14.6 | 12.5 | 8.3 | 7.8 | 7.3 | 6.9 |
| Pre-Mandate Removal Average from Jan 27 '21 | 7.0 | 6.5 | 4.5 | 4.8 | 4.2 | 3.8 |
| Post-Mandate Removal 6 week Average | 3.5 | 5.5 | 5.4 | 1.7 | 2.1 | 2.5 |
| Percent Change from Sep 16 '21 vs Pre-Removal | -60.8% | -30.7% | -2.4% | -64.8% | -55.2% | -45.1% |
| Percent Change from Dec 2 '21 vs Pre-Removal | -75.8% | -56.3% | -35.2% | -78.1% | -71.3% | -63.9% |
| Percent Change from Jan 27 '21 vs Pre-Removal | -49.6% | -15.9% | 19.1% | -64.7% | -50.2% | -33.8% |
| Post Mandate Removal 12 week average | 7.0 | 12.0 | 8.2 | 4.3 | 4.6 | 5.0 |
| Percent Change from Sep 16 '21 vs Pre-Removal | -22.0% | 52.9% | 49.1% | -12.4% | -2.3% | 9.8% |
| Percent Change from Dec 2 '21 vs Pre-Removal | -51.8% | -3.5% | -1.0% | -45.5% | -37.4% | -27.8% |
| Percent Change from Jan 27 '21 vs Pre-Removal | 0.2% | 85.5% | 82.0% | -12.2% | 8.4% | 32.2% |

**Table 2** presents differences in case-rates in staff and students before and after March 3rd, as well as the difference in these differences, between (i) the treatment group (Boston-Chelsea); (ii) 70 other districts in Greater Boston used as controls by Cowger et al, and(iii) a larger control group of 217 districts statewide. We also present the ratio of case-rates in each group, as an alternative to the difference. The DiD coefficient is positive in three of the six scenarios, indicating that cases actually rose by a larger degree in the districts maintaining the mask mandates than the controls. In all scenarios, including the three where the coefficient is negative, the effect is not statistically significant.

**Table 2:** DiD analysis between different "treatment groups" by testing population

| Panel A: (Total Student Cases in Group)/(Total Enrollment in Group, in 000s) | | | | |
|---|---|---|---|---|
| | Sep 21--Feb 22 | Mar 22--June 22 | Enrollment (000s) | |
| 1. Boston and Chelsea | 4.86 | 4.21 | 52.2 | |
| 2. Greater Boston districts (70) | 8.3 | 7.57 | 241.8 | |
| 3. Other MA districts (217) | 9.22 | 4.94 | 534.2 | |
| | | | Diff-in-diff | p-value |
| Difference of 1 and 2 | -3.44 | -3.36 | 0.08 | 0.982 |
| Difference of 1 and 3 | -4.36 | -0.73 | 3.63 | 0.291 |
| Ratio of 1 to 2 | 0.59 | 0.56 | | |
| Ratio of 1 to 3 | 0.53 | 0.85 | | |

| Panel B: (Total Staff Cases in Group)/(Total Staff FTE in Group, in 000s) | | | | |
|---|---|---|---|---|
| | Sep 21--Feb 22 | Mar 22--June 22 | Staff FTE (000s) | |
| 1. Boston and Chelsea | 11.5 | 7.27 | 9.4 | |
| 2. Greater Boston districts (70) | 9.29 | 12.31 | 37.2 | |
| 3. Other MA districts (217) | 11.04 | 10.53 | 82.4 | |
| | | | Diff-in-diff | p-value |
| Difference of 1 and 2 | 2.21 | -5.04 | -7.25 | 0.211 |
| Difference of 1 and 3 | 0.46 | -3.26 | -3.72 | 0.529 |
| Ratio of 1 to 2 | 1.24 | 0.59 | | |
| Ratio of 1 to 3 | 1.04 | 0.69 | | |

| Panel C: (Student+Staff Cases in Group)/(Enrollment+Staff FTE in Group, in 000s) | | | | |
|---|---|---|---|---|
| | Sep 21--Feb 22 | Mar 22--June 22 | Staff FTE (000s) | |
| 1. Boston and Chelsea | 5.87 | 4.67 | 61.6 | |
| 2. Greater Boston districts (70) | 8.44 | 8.2 | 279 | |
| 3. Other MA districts (217) | 9.46 | 5.69 | 616.6 | |
| | | | Diff-in-diff | p-value |
| Difference of 1 and 2 | -2.57 | -3.53 | -0.96 | 0.802 |
| Difference of 1 and 3 | -3.59 | -1.02 | 2.57 | 0.481 |
| Ratio of 1 to 2 | 0.7 | 0.57 | | |
| Ratio of 1 to 3 | 0.62 | 0.82 | | |

**Sensitivity Analysis**

We also performed a sensitivity analysis comparing the two major Omicron waves during 2021/2022. In this analysis, the ratio of cases in Boston-Chelsea to the 70 unmasked districts was *higher* in spring than in winter, moving from 118% lower in Boston-Chelsea than the other 70 districts during the first wave when all districts had mask mandates, to only 46.9% lower during the Spring wave when only Boston-Chelsea maintained mask mandates (**Supplementary Material**). Additionally, we found that Suffolk County, which includes Boston-Chelsea, consistently reported lower case-rats throughout 2021/22 compared to the 12 other counties in Massachusetts (**Figure S2 in Supplementary Material**).

**Testing Rates**

As per MDESE,[6] consent to SARS-CoV-2 testing in early 2022 was higher (70% vs 58%) among more-vaccinated schools (>80% vaccination rate) than those with lower vaccination rates (<50%). This is another clear time-varying confounder precluding DiD use for causal inference.

**Community Prior Immunity**

Prior community SARS-CoV-2 infection (PI) burden was highest in Boston-Chelsea (**Figure S3 in the Supplementary Appendix**). We identified a moderate to strong relationship ($R^2 = 0.35$-$0.66$; p-values <0.001) between prior community infection burden and district case rates in Spring 2022, with prior immunity alone explaining as much as two-thirds of the variation in case rates in Spring of 2022. (**Figure 3**). Furthermore, the inverse relationship between infection rates post-mandate versus pre-mandate infection rates demonstrated a dose-response relationship (**Figure 3**): a larger level of pre-mandate infections resulted in a larger decrease in post-mandate infections.

**Figure 3:** Analysis of the relationship between community case rates from 12/1/2020 through 2/27/2022 and case rates post 2/27/2022.

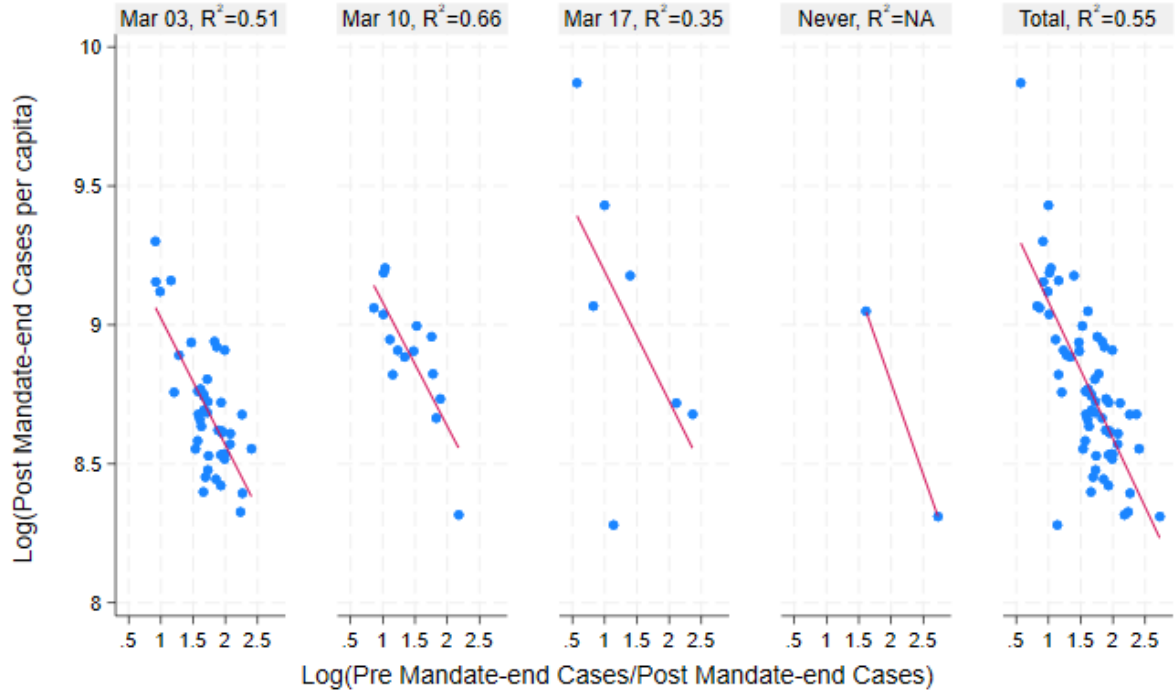

| Dependent Variable: Post-2/27/2022 Community Cases by District per 100k Population | | | | |
|---|---|---|---|---|
| | Regression | | ANOVA | |
| Independent Variable | R2 | Root Mean Square Error | F | Significance F |
| Pre-2/27/2022 Cases (Omicron only) over Post-2/27/2022 Cases per 100k | 0.55 | 0.20 | 79.28 | **p<0.001** |
| March 3rd End Only (n=44) | 0.51 | 0.16 | 42.92 | **p<0.001** |
| March 10th End Only (n=15) | 0.66 | 0.13 | 24.84 | **p<0.001** |
| March 17th End Only (n=7) | 0.35 | 0.47 | 2.66 | **p<0.001** |

Legend: Negative relationship between prior community infection burden and subsequent case rates including correlation coefficients by districts drop dates or no drop (top). Significance testing on correlation coefficients (bottom).

**Discussion**

We re-analyzed and expanded the analysis by Cowger et al using classic difference in differences and a statewide dataset and failed to find evidence of causal or significant link between mask mandate policies and district case rates. Differences in case rates before and after-March 3rd, 2022, in the districts that maintained mask mandates remained similar compared to the 70 other districts in the Greater Boston Area that dropped mask mandates. Maintaining mask mandates was associated with a non-significant increase in case rates relative to the rest of the districts in the state.

Compared with Cowger et al, our analysis found a drop in cases in the Boston and Chelsea districts which was smaller in magnitude, non-significant and not robust to either 1) comparison to the state at large, or 2) the individual districts stratified by length of time they dropped the mandates.

Our results differ from Cowger et al for at least four reasons. First, we compare case-rates across groups normalized by population, rather than the average within each district group. Second, we did not employ the Callaway Sant'Anna DiD methodology as we believe this was inappropriate in the context of a few weeks' difference in mandate-lifting and lack of dose-response relationship with lifting date. Third, we may have found smaller magnitude and non-significant differences because Cowger et al stated their "difference-in-differences estimates were essentially zero" prior to March 3rd and it is possible that the authors truly considered the differences to be zero prior to the mask policy changes in their analyses; Cowger et al. did not provide transformed data and codes for verification. Fourth, we included a statewide analysis and, unlike Cowger et al,[5] did not exclude districts which reported no cases for >5-10 weeks.

There are also at least five reasons why Cowger et al's causal inference was inappropriate. First, they limited their control group to 70 districts without any explanation and, when we extended the analysis to all districts statewide, case-rates were non-significantly higher in Boston-Chelsea, which continued to mandate masks.

Second, their results were not robust to length of time under policy, even showing an inverse relationship between districts dropping their mandates and case-rates post-mandate. This argues against a causal relationship between mask mandates and case-rates, rendering the Callaway Sant'Anna method of using DiD to assess the effects of the staggered mask drop dates a methodological choice that is difficult to defend, particularly without adjusting for community rates. We used a single timepoint for removal of mask mandates and found that Boston-Chelsea had the lowest case-rates throughout the academic year, with districts that dropped mask mandates first having the largest relative drop in case-rates.

Third, the Boston district was an outlier (**Figure 1**) in terms of size, demographics, and racial composition thereby amplifying the impact of idiosyncrasies specific to this district. The CDC, for example, reported lower testing rates among black and lower income Americans.[15] Lower SEC families are also less likely to purchase at-home tests or report positive results.

Fourth, pursuant to the above, time-varying confounders precluded the use of DiD for causal inference, including PI, testing and reporting rates, and community disease. Additionally, mask mandates are invariably accompanied by other mitigations, which will significantly confound studies assessing masking impact only. Likewise, changes in masking policies will invariably be accompanied by other policy and behavioral changes, especially in educational settings.[12] Indeed, according to MDESE, testing practices changed substantially when mask mandates were removed.

Finally, we found an inverse dose-response relationship between PI before and case-rates after removal of mask mandates. This is consistent with a strong protective effect of prior infection against reinfection (**Supplementary Material**) and highlights why it is problematic to assume that because the Boston-Chelsea had previously been "hit hardest by COVID-19"[5] that they would continue to be in terms of case rates, given these districts were likely most protected from reinfection because of existing infectious immunity.

The strengths of our analysis include expanding the comparison group to the entire state providing a wider picture of how Boston-Chelsea case-rates changes throughout the academic year, including a sensitivity analysis, and providing an analysis of PI. We also demonstrate why both the choice of DiD methodology and the specific Santa'Anna method, which looks at staggered policy change effects, were inappropriate in previous analysis.[5]

Our analysis, like Cowger et al, was limited by not having full access to testing data, seroprevalence estimates of population immunity, or whether the cases came from school or the community. Additionally, because of the retrospective observational methodology, we cannot draw definitive conclusions about mask mandates because of so many other potential confounders. Cluster randomized trials would be preferable, but have yet to be performed in an educational setting. Furthermore, focusing on Massachusetts is itself arbitrary, but our aim was a comparative re-analysis of a study that likely influenced and continues to influence local and national policy. Considering the large variation in school masking practices globally, there is an enormous number of datasets which could be similarly analyzed. This large analytic flexibility suggests that all conclusions are possible using observational data.[12,34]

Selective restriction of data by time and/or geography can significantly affect outcomes as we have previously demonstrated with mask policies.[35] Cowger et al's conclusions on mask effectiveness against COVID-19 in schools – and in particular, their unrealistically high protective estimates of masking compared to other studies – are at odds with higher quality evidence for COVID-19.[1,2]

**Conclusion**

For multiple reasons, the results of Cowger et al do not hold up to re-analysis. We failed to identify evidence of a causal relationship, or even consistent association, between mask mandates and district SARS-CoV-2 infection rates. The findings of Cowger et al should not be used as evidence mask mandates prevent the spread of SARS-CoV-2 in educational settings, nor, as the authors suggested, that they may be useful for mitigating the effects of structural racism.

**Acknowledgements**

The authors report no funding for this study.

**Declaration of interests**

The authors report no relevant conflicts of interest.

# SUPPLEMENTARY APPENDIX

**Extension of study to all districts state-wide**

The Massachusetts statewide school mask mandate was dropped February 28th, 2022. For the 289 Massachusetts districts that met Cowger et al's inclusion criteria, we performed an extensive Google search for news articles on the date of ending mask mandates.[1] Various media tracked these districts. We searched for articles in Massachusetts media that reported dates of dropping, retaining or reinstating mask mandates.[1,2,3,4] Other news sources are available from the authors upon request. We identified three districts that kept mask mandates in place beyond March 17: Northampton, Springfield and Amherst-Pelham. All three districts ended these rules by March 31. However, at least two school districts reinstated masks in May: Brookline and Northampton. Overall, Boston and Chelsea were the only two districts in the state that maintained mask rules continuously for the 2021-22 academic year. While 5–7 districts had sporadic mask rules in place beyond March 17, the vast majority of the state's 289 districts did not.

In some specifications, we analyze data from the CDC which are at the county, rather than district, level. To do this we compare reported case rates for Suffolk County, which contains Boston and Chelsea, with all other counties in the state, where mask rules were almost uniformly dropped by March 17.

**Sensitivity analysis of major infectious waves**

During the height of the Omicron wave in January 2022, the three week moving average of cases per 1000 students was 118.9% lower in Boston and Chelsea versus the other districts (13.2/1000 vs 28.9/1000 students, respectively). The corresponding figures at the peak of the Spring wave in May were only 46.9% lower in Boston and Chelsea (9.2 and 17.0/1000, respectively). Thus the relative case rates in Boston and Chelsea *were higher* in comparison to the other districts (moving from 118% lower to only 46.9% lower) while they retained mask requirements during the spring wave, which is again inconsistent with the conclusion drawn by Cowger et al.

---

[1] Various media tracked these districts. We searched for articles in Massachusetts media that reported dates of dropping, retaining or reinstating mask mandates. Sources include [23], [24], [25] and [26]. Other news sources are available from the authors upon request.

**Community case rates by Massachusetts district over the 2021-2022 school year**

During December–February, when all school districts in the state required masks, Boston and Chelsea reported average weekly cases per capita that were generally 50–55% of the average rates reported in the rest of the state. During April-June, Boston and Chelsea's average case rate was over 70% of the rest of the state. In other words, though the Boston and Chelsea districts maintained their mask mandates, they had relatively higher rates of infection (going from 50-55% of the comparison group to 70% after the comparison group lifted mask mandates) after the other districts lifted mask mandates than when all (or almost all) districts were requiring masks.

From September 2021 until March 2022, Suffolk County consistently ranked as the lowest or second-lowest of the 13 counties that we analyzed, according to weekly student cases per capita. It was in fact only during the period studied by Cowger et al that this changed: Suffolk ranked between seventh and ninth among the 13 counties during April and May 2022, even though it was the only county with widespread school mask mandates during this time.

In May, Boston and Chelsea had higher average weekly case rates than the average of the rest of the state, even though they were the only ones to require masks throughout. Overall, this state-wide comparison is evidence against the claim that mask mandates were responsible for lower case rates in Boston and Chelsea.

**Regression methods for analysis of prior community immunity**

We performed univariate regression of the log proportion of infections after the end of statewide mandates on the proportion of total infections that occurred prior to the end of the mandate, in order to assess the impact of prior community infection burden. We also estimated multiple regression models of per-capita case rates in each group of districts on an indicator variable for the treatment group, an indicator for post-March 3 observations, as well as the interaction of these two variables. The p-values reported in **Figure 3** of the paper correspond to the estimated coefficient on these interaction terms.

**Assessment of correlation of post mandate change cases with prior community immunity levels**

The authors appeared to have collected staff and student case rates as a percentage of total Omicron cases through February 27$^{th}$, 2022 (Table S7 page 30 of Supplement) but did not appear to perform an analysis assessing correlation with this variable. Notably, in the data the authors presented, there was a trend towards pre-February 28$^{th}$ district infection burden negatively correlating with post-February 27th infection burden.

We found community COVID-19 infection (PI) burden was highest in the group that continued masking prior to the termination of the statewide mask mandate (Fig S2).

**Figure S3** shows cumulative community case rates separately for the four groups of districts, using data from the Massachusetts Department of Elementary and Secondary Education. The black curve represents cumulative cases for the two districts that maintained mask mandates throughout the year which, as we have already pointed out, is dominated by the City of Boston. This group had the highest reported cumulative cases per capita throughout the study period. All four groups experienced a rapid increase in cases during December 2021-January 2022. By the time that mask mandates were mostly dropped in early March, Boston had significantly higher cumulative case burden than other districts. This is additional evidence that prior infections in the community during the Omicron wave are likely to have led to lower susceptibility to infections in the Spring of 2022 following the end of school mask requirements.

Furthermore, post-March 3rd infection burden varied with pre-March 3rd infection burden in a "dose-response fashion" (**Figure 3**, **Figure S3**) where the amount of post-mandate infection burden had a significant negative relationship with the pre-mandate infection monotonically and not the mandate lift date (i.e., if masking policy change was the driver, then we would expect an earlier mandate end date to monotonically lead to higher post-mandate case burden). Infection-based immunity increased the most during the early omicron period in the district that continued their mask mandate and had the lowest infection burden during the study period.

**Figure 3** shows the relationship between infections before and after the end of statewide mask mandates, separately for each of the four groups of districts as well as combined for all 68 districts. Each panel shows a scatter plot of case rates after the end of mask mandates against the ratio of pre-mandate to post-mandate cases, as well as the line of best fit. In all groups, as well as the overall set of districts, there is a clear negative relationship, suggesting that areas with a higher burden of infection prior to the end of mask-mandates had lower post-mandate cases. We verify this using a different data source from the Centers for Disease Control (CDC) and show this in **Figure S4**. The CDC reports data at the county level, which is more aggregated than the city and town level data from Massachusetts. However, data reported by the CDC must conform to strict standards, including patient level information on demographics and outcomes, which is more rigorous than the aggregate counts of cases available from other sources. **Figure**

**S4** reports cumulative case counts for three Massachusetts counties. Although school districts do not always map cleanly to counties, the three counties presented here correlate very well with districts that dropped or maintained mask requirements. Over 90% of Suffolk county's population lies in the two districts of Boston and Chelsea that maintained mask mandates. All districts in Norfolk and Middlesex counties dropped mask requirements in schools, and these two counties account for 52 out of the 66 districts that did so.

The differences in case rates reported in the Cowger et al can be better explained by pre-existing immunity and differences in testing rates than changes in district mask policy. Existing omicron-based community immunity would be expected to influence the student and staff case rates found during the study period and we have demonstrated a dose-response relationship with prior immunity that was not seen with mask mandate lifting date.

We have compelling evidence that prior infection confers significant, albeit not absolute, resistance to reinfection. A prospective 12-month antibody kinetics study of 351 Italian children plotted SARS-CoV-2 anti-RBD antibodies longitudinally, stratified by age.[5] There are also some differences related to children in particular. A German household study followed a large cohort of children for a year to measure the humoral responses to S and N proteins in adults and children.[6] At 3-4 months, the children had significantly higher antibody titers than adults against the spike protein, the receptor binding domain (RBD), and the S1 and the N proteins ($p<0.001$). At 12 months, only 3.8% of children had sero-reverted compared to 17.1% of adults. In contrast, antibody decay over 4-6 months following mRNA vaccination is well established and manifests in clinical outcomes, such as symptomatic COVID-19.[7-13] Higher levels of pre-existing immunity in districts would be expected to lead to lower case rates during the study period. We have compelling evidence that prior infection confers significant, albeit not absolute, resistance to reinfection, and that this response is durable for at least 12 months while vaccine efficacy against the same substantially wanes in 4-6 months.[7-20] Infection acquired immunity provides symptomatic protection even against variant lineages with a high degree of immune escape like omicron. Vaccine effectiveness against any symptomatic omicron infection was higher among those infected more than a year prior compared to those boosted at least a month prior.[21] Our analysis demonstrates a dose-response relationship with prior infection-related immunity that was not seen with the mask mandate lifting date as plotted in Cowger et al. (Fig 1d). The dose-response relationship suggests a causal link beyond simple correlation.

On the cellular immunity side of the equation, Dowell et al., found that spike-specific T cell responses were higher in children than adults and sustained for at least 12 months, even in children without detectable antibodies to SARS-CoV-2.[26] Supporting the separate impact of cellular immunity, a study of pre-pandemic cellular and humoral immunity to common-cold coronaviruses (CCC) and other respiratory pathogens, Yu, Sette and colleagues found that "high CD4+ T cell reactivity to HCoV, but not antibody responses, was associated with high pre-existing SARS-CoV-2 immunity."[15] Vaccines induce a strong and reliable serological response, but are less effective at eliciting antibodies in the mucosa and in the lung and respiratory tissues.[28,29] These tissue-resident immune responses – both humoral and cellular components – are critical for preventing infection and disease progression.[18] A recent study found that despite neutralization escape against Omicron, CD4 and CD8 T cells retained 70-80% cross-reactivity against conserved epitopes on S, M and N proteins.[19] These cross-reactive T cells, derived from vaccination or infection, are thought to play an important role in limiting progression to severe disease. Those with a history of infection (with or without vaccination) have developed both humoral and cellular recognition of the membrane and nucleocapsid proteins in addition to spike protein.[20,21] Infection acquired immunity provides symptomatic protection even against variant lineages with a high degree of immune escape like omicron. Vaccine effectiveness against any symptomatic omicron infection was higher among those infected more than a year prior compared to those boosted at least a month prior.

The differences in case rates reported in the Cowger et al can be better explained by pre-existing immunity and differences in testing rates than changes in district mask policy. Existing omicron-based community immunity would be expected to influence the student and staff case rates found during the study period and we have demonstrated a dose-response relationship with prior immunity that was not seen with mask mandate lifting date.

Given the above, as immunity levels were changing throughout the winter and into the spring study period, it was inappropriate for Cowger et al to use a DiD analysis to infer causality given no correction for time-varying confounding variables. Consequently, the study should be viewed as simply as any other observational study of masking.

Differences in testing rates owing to discrepancies in testing consent and a shift to more at-home testing also both precluded the use of DiD methodology by Cowger et al and may have led to lower identified COVID-19 cases during the spring of 2022 study period in the Boston and Chelsea districts.

**Reasons for discrepancies between our results and those of Cowger et al.**

**Table 1** shows an analysis of case rates in three different groups: (i) the two districts, Boston and Chelsea, that maintained school mask mandates continuously, (ii) the 70 other districts in the Greater Boston area studied by Cowger et al, all of which dropped school mask mandates between reporting weeks March 3 and March 17 and (iii) 217 other districts in Massachusetts, virtually all of which also dropped mask mandates by March 17, but which were not studied by Cowger et al.

The results of **Table 1** show the difference in the differences pre- and post-March 3rd 2022 for the treatment group, consisting of group (i) above, compared with each of the two control groups, defined as (ii) and (iii) respectively. The results do not show a statistically significant relationship between school mask mandates and per capita infection rates, in contrast to the findings of Cowger et al.

There are at least four reasons for the discrepancy between our results and those of Cowger et al. (a). **Table 1** presents aggregate cases among students and staff for each group by total student/staff population; separated into two time periods (before/after March 3rd, 2022). By contrast, the regression specifications used by Cowger et al use each school district as a distinct observation. We argue that this is inappropriate in the current context as the Boston school district is the vast majority of the treatment group described in (i) above. We believe it is more appropriate to compare case rates across groups normalized by population, rather than the average, within each district group. The latter is the method chosen by Cowger et al, and produces a spuriously high burden of cases in the post-mandate treatment period.

(b). Cowger at al use other covariates in their regressions, as well as the Callaway and Sant'anna (2021) DiD method to exploit the staggered dropping of mask mandates across districts. We do not implement these modifications to the regressions presented in Table 1. We do not believe it is appropriate to estimate the multiple time period analysis of Callaway and Sant'Anna (2021) in this context as the two-week difference in dropping mask mandates does not provide meaningful variation and is likely to introduce unnecessary noise given the likelihood of concomitant policy or behavior changes at the time the mask mandates were dropped. Additionally, as shown in the article, we find no evidence that districts that were first to drop mask mandates had faster increases in cases; in fact the opposite was true, which further undermines the use of the staggered DiD approach. However, we have made our raw data available for re-analysis if other groups wish to replicate the Callaway and Sant'Anna employed by Cowger et al. The first-order variation is provided by the February 28 end of the statewide mask mandate, which were quickly dropped by most districts, but which were retained for over 20 more weeks by two districts in particular.

(c) We also perform a second, expanded analysis, using a much larger control group, consisting of 217 districts, which Cowger et al do not use. As explained in the paper, we believe their restriction of analysis to the Greater Boston area was arbitrary, as the dropping of statewide mandates provided an opportunity to exploit district level variation in almost 300 school districts.

The results using the wider set of districts in the statewide control group do not support the notion that cases rose more slowly in the treatment group that retained mask mandates.

(d) When we performed the statewide analysis we did not exclude districts which reported no to low cases over 5-10 week periods as done by Cowger et al.

**Possible incorrect dates of districts' mask policy changes**

We have been notified by anonymous personal communication that 24 schools had already lifted their mask mandates prior to 2/28/2022, which was the first mask-mandate lifting date per Cowger, et al. Thus the allegation from a source familiar with the data is that schools were incorrectly classified as lifting mask mandates on or after 2/28/2022. If the claim is true, this would preclude the use of the DiD technique and necessitate consideration of all possible confounding differences between the masked and unmasked districts during the study period.

**Analysis of district cases by county**

We verify this using a different data source from the Centers for Disease Control (CDC) and show this in **Figure S4**. The CDC reports data at the county level, which is more aggregated than the city and town level data from Massachusetts. However, data reported by the CDC must conform to strict standards, including patient level information on demographics and outcomes, which is more rigorous than the aggregate counts of cases available from other sources. The figure above reports cumulative case counts for three Massachusetts counties. Although school districts do not always map cleanly to counties, the three counties presented here correlate very well with districts that dropped or maintained mask requirements. Over 90% of Suffolk county's population lies in the two districts of Boston and Chelsea that maintained mask mandates. All districts in Norfolk and Middlesex counties dropped mask requirements in schools, and these two counties account for 52 out of the 66 districts that did so.

SUPPLEMENTARY FIGURES

**Figure S1.** Extension of the time period of figure 2 but with students and staff combined (L) and staff alone (R)

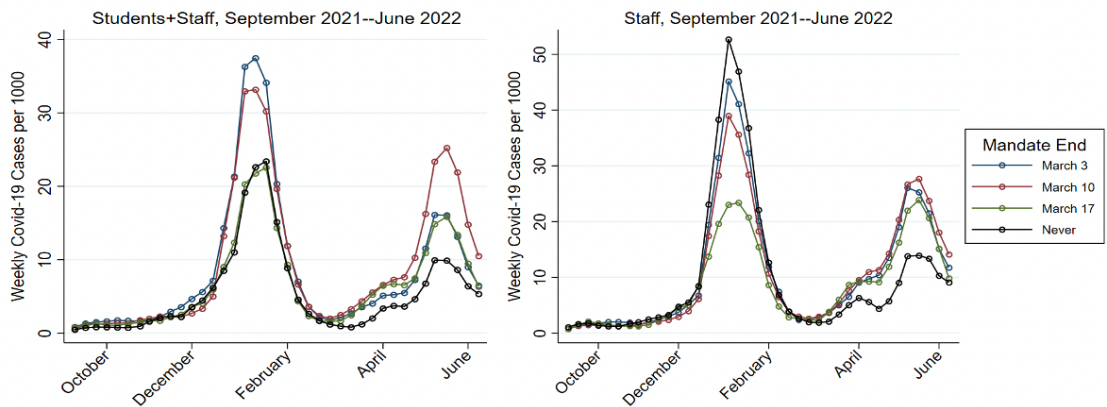

**Figure S2:** Student case rates reported by districts by county with Boston and Chelsea in Suffolk County shown in red.

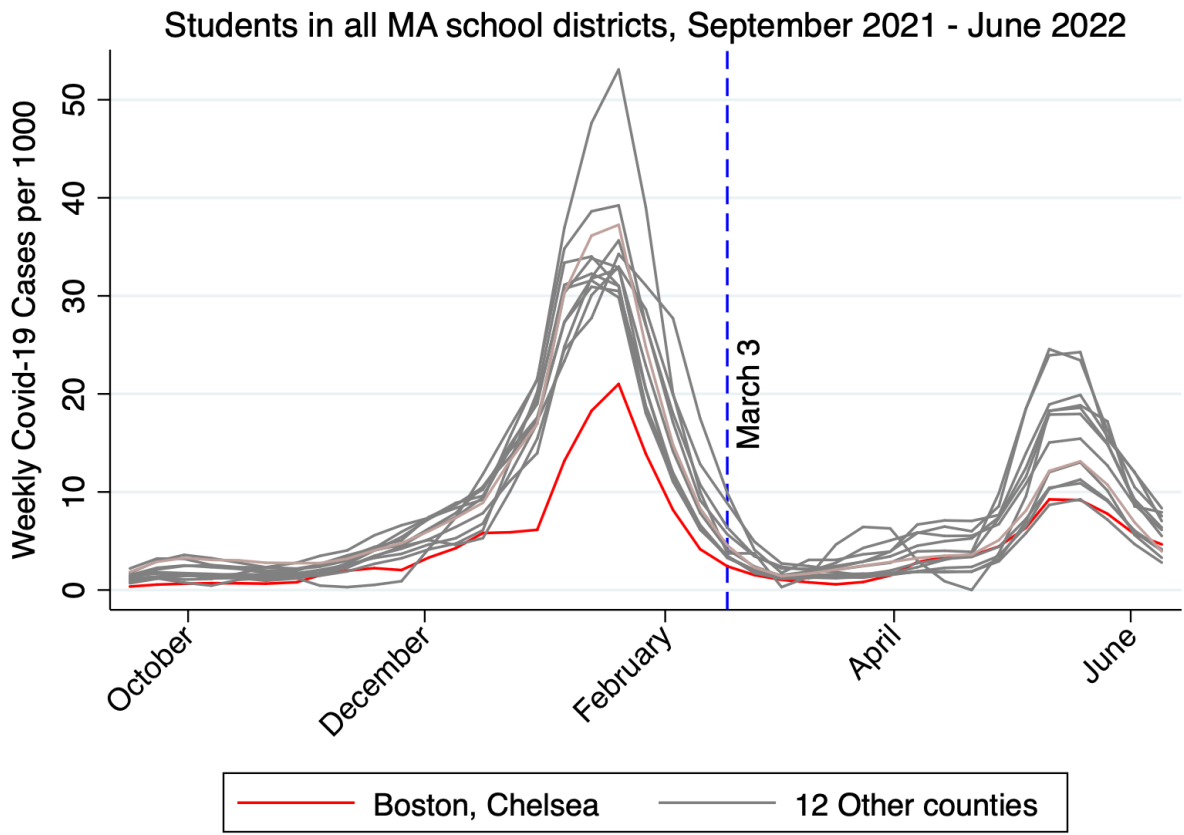

**Figure S3.** Cumulative community cases by district mask policy

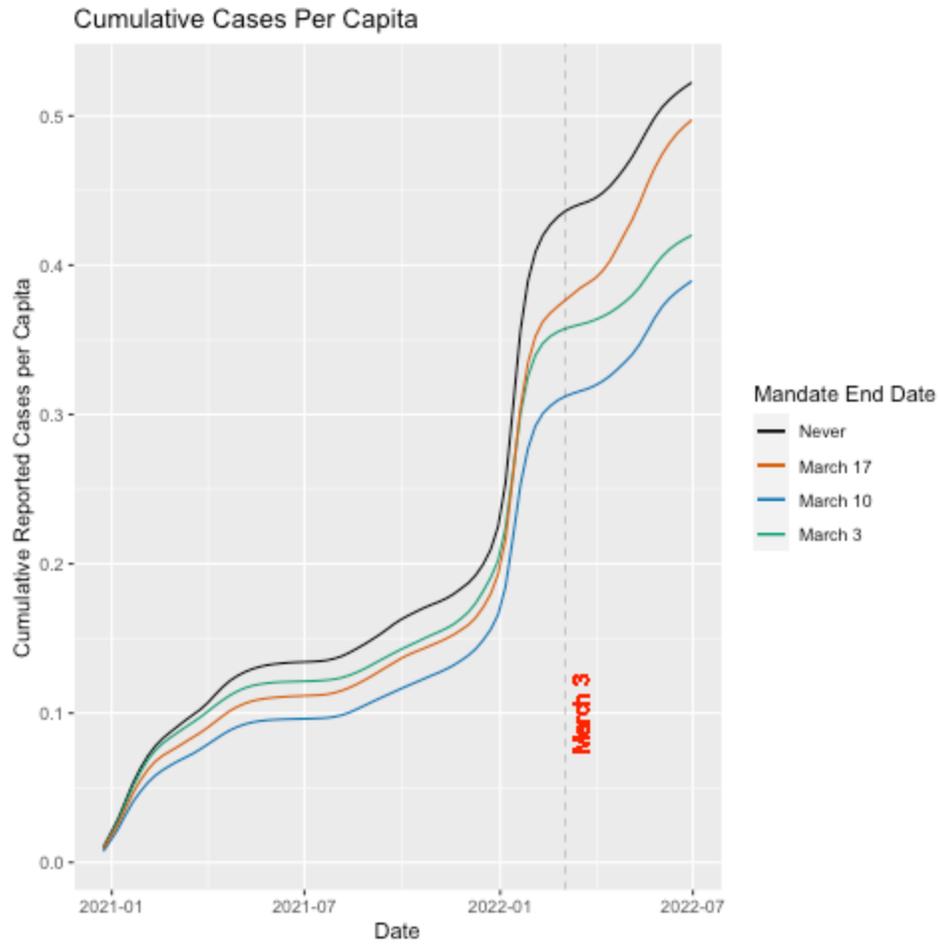

**Figure S4**. Cumulative cases per capita by county included in Cowger et al.

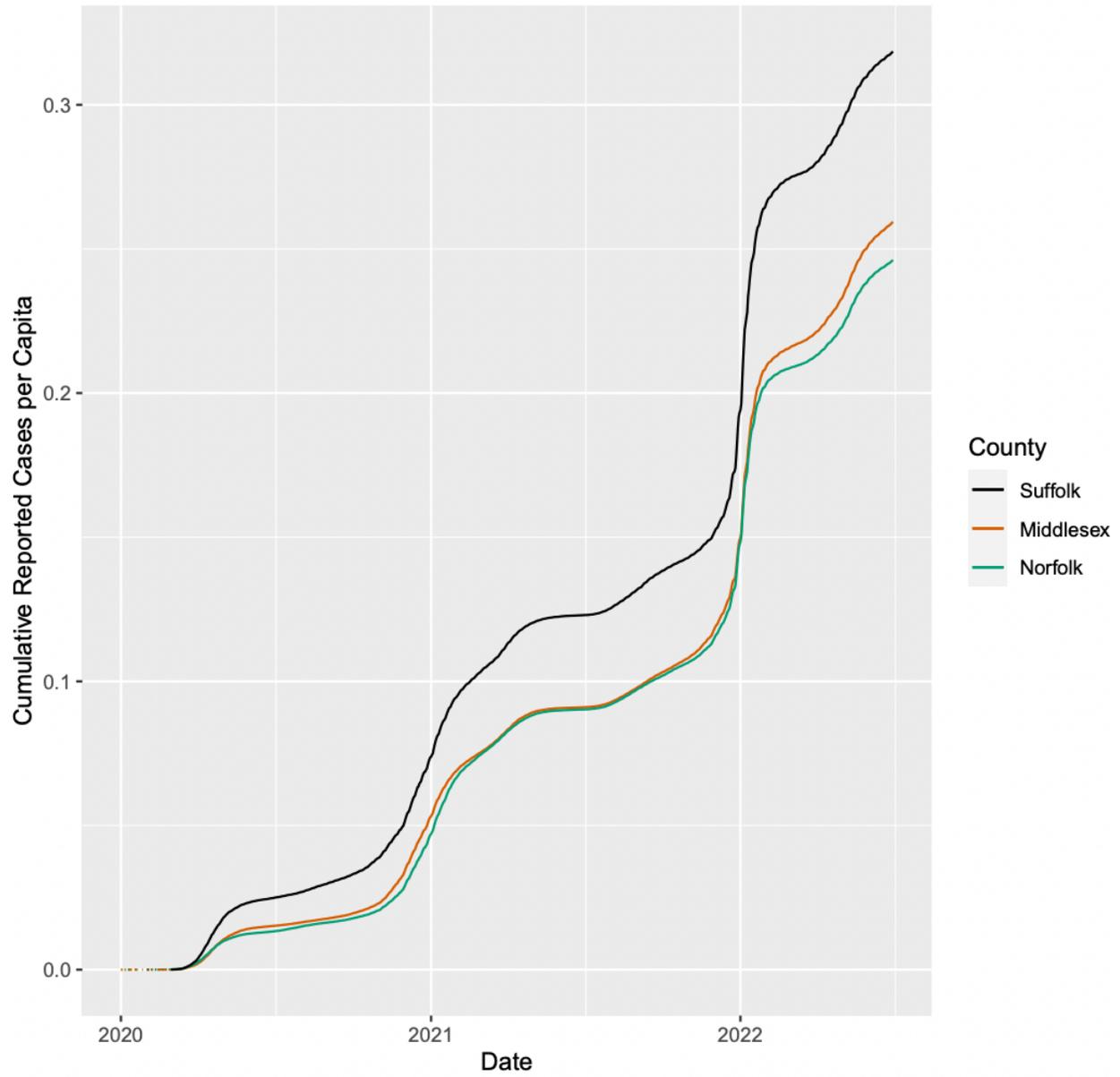